\documentclass[lettersize,journal]{IEEEtran}

\usepackage[rgb]{xcolor}
\selectcolormodel{natural}
\usepackage{ninecolors}
\usepackage{tudacolors}
\usepackage{xspace}

\colorlet{accent}{TUDa-10c}
\colorlet{accent2}{azure4}
\colorlet{accent3}{cyan4}
\newcommand{\accentname}{purple\xspace}
\newcommand{\accenttwoname}{blue\xspace}
\newcommand{\accentthreename}{teal\xspace}

\definecolor{colorTh}{RGB}{160, 20, 0}
\definecolor{colorFWgrid}{RGB}{160, 20, 0}

\colorlet{colorP1}{black}
\colorlet{colorP2}{TUDa-10c}
\colorlet{colorP3}{TUDa-1c}
\colorlet{colorP4}{TUDa-4b}
\colorlet{colorP5}{TUDa-6b}
\colorlet{colorP6}{TUDa-8b}
\definecolor{colorP7}{RGB}{0, 100, 0}
\definecolor{colorP8}{RGB}{0, 128, 255}
\definecolor{colorP9}{RGB}{0, 0, 0}
\definecolor{colorOne}{RGB}{160, 20, 0}
\definecolor{colorTwo}{RGB}{0,0,255}
\definecolor{colorThree}{RGB}{0,0,0}

\colorlet{tablelinecolor}{gray6}
\colorlet{tablezebracolor}{gray9!50!white}

\usepackage[english]{babel}
\usepackage[babel, autostyle]{csquotes}
\usepackage{xparse}	
\usepackage{ifthen}
\usepackage{tabularray}
\UseTblrLibrary{booktabs}
\usepackage{colortbl}
\SetTblrInner{rowsep=2pt,colsep=4pt} 
\usepackage{acro}
\usepackage[mode=image]{standalone}
\usepackage{subcaption}
\usepackage{import}
\usepackage{upgreek}
\usepackage{textcomp}
\usepackage{xspace}
\usepackage{todonotes}
\usepackage{textcomp}

\usepackage{mathtools}
\usepackage{amssymb}
\usepackage{amsmath}
\allowdisplaybreaks
\usepackage{siunitx}
\usepackage{nicefrac}

\DeclareSIUnit\epszero{\permittivityvacuum}
\DeclareSIUnit\muzero{\permeabilityvacuum}
\DeclareSIUnit\nuzero{\reluctivityvacuum}

\usepackage{tikz}
\usepackage{pgfplots}
\usepackage{pgfplotstable}
\pgfplotsset{compat=1.18}
\usepackage{pgffor}
\usepackage{pgfmath}
\usepackage{circuitikz}
\usetikzlibrary{decorations.pathmorphing, decorations.pathreplacing, decorations.shapes,decorations.markings,math,3d,calc,arrows.meta, positioning, pgfplots.colormaps,patterns, patterns.meta}
\usepgfplotslibrary{colorbrewer, patchplots}
\tikzset{>=Stealth}
\tikzset{level/.style={%
		execute at begin scope={\pgfonlayer{#1}},
		execute at end scope={\endpgfonlayer}
}}
\pgfplotscreateplotcyclelist{myColorCycleList}{black,TUDa-2b,TUDa-9b,TUDa-3c,TUDa-7b,TUDa-11b,TUDa-10c,TUDa-8c}
\pgfplotsset{cycle list name={myColorCycleList}}
\pgfplotsset{every axis plot/.style={thick,mark=none}}
\pgfplotsset{field plot element/.style={line width=0pt,faceted color=none}}
\pgfplotsset{field plot node/.style={line width=0pt,faceted color=none,shader=interp}}

\graphicspath{{figures}}

\usepackage{hyperref}
\hypersetup{breaklinks=true,colorlinks=true,citecolor=blue,urlcolor=blue}

\usepackage{fancyhdr}
\newcommand{\Pyrit}{\emph{Pyrit}\xspace}

\newcommand{\ComsolMultiphysics}{\textsc{Comsol}\xspace Multiphysics\textsuperscript{\textregistered}\xspace}
\newcommand{\Python}{Python\xspace}

\newcommand{\transpose}{^{\mkern-1.5mu\mathsf{T}}}
\newcommand{\setdelimiter}{:}
\newcommand{\setprop}[2]{\left\{#1\setdelimiter#2\right\}}
\newcommand{\R}{\mathbb{R}}
\newcommand{\N}{\mathbb{N}}

\newcommand{\frequency}{f}
\newcommand{\angularfrequency}{\omega}
\newcommand{\skindepth}{\delta}
\newcommand{\permeabilityvacuum}{\mu_0}
\newcommand{\jomega}{\jmath\angularfrequency}

\newcommand{\conductivity}{\sigma}
\newcommand{\conductivityconductor}{\conductivity_{\mathrm{c}}}
\newcommand{\conductivityinsulator}{\conductivity_{\mathrm{i}}}
\newcommand{\reluctivity}{\nu}
\newcommand{\reluctivityconductor}{\reluctivity_{\mathrm{c}}}
\newcommand{\reluctivityinsulator}{\reluctivity_{\mathrm{i}}}
\newcommand{\permeability}{\mu}
\newcommand{\permeabilityconductor}{\permeability_{\mathrm{c}}}
\newcommand{\thermalconductivity}{\kappa}
\newcommand{\thermalconductivityconductor}{\thermalconductivity_{\mathrm{c}}}
\newcommand{\thermalconductivityinsulator}{\thermalconductivity_{\mathrm{i}}}

\newcommand{\thConv}{h}

\newcommand{\dS}{\diff S}

\newcommand{\intsep}{\,}

\newcommand{\qmatmg}{\mathbf{q}_\mathrm{mg}}
\newcommand{\qmatth}{\mathbf{q}_\mathrm{th}}
\newcommand{\sourcecurrentdensity}{\vec{J}_{\mathrm{s}}}
\newcommand{\domainboundary}{\Gamma}

\newcommand{\heatsrc}{q }
\newcommand{\heatsource}{\heatsrc_\mathrm{v}}

\newcommand{\cV}{c_\mathrm{v}}

\newcommand{\volheatcap}{c_\mathrm{v}}

\newcommand{\temperature}{T}
\newcommand{\temperatureRef}{\temperature_{\mathrm{ref}}}
\newcommand{\temperatureAmb}{\temperature_{\mathrm{amb}}}
\newcommand{\conductivityRef}{\conductivity_{\mathrm{ref}}}
\newcommand{\temperaturecoefficientRef}{\alpha_{\mathrm{ref}}}
\newcommand{\temperatureFE}{\vartheta}
\newcommand{\temperaturemat}{\boldsymbol{\temperatureFE}}

\newcommand{\energythhom}{U_\mathrm{hom}}
\newcommand{\energythres}{U_\mathrm{res}}

\newcommand{\mvp}{\vec{A}}
\newcommand{\mvpFEmat}{\mathbf{a}}
\newcommand{\mvpFE}{a}

\newcommand{\scalarpot}{\phi}

\newcommand{\turnwidth}{b}
\newcommand{\turnwidthc}{\turnwidth_\mathrm{c}}
\newcommand{\turnwidthi}{\turnwidth_\mathrm{i}}
\newcommand{\foilwidth}{w}
\newcommand{\foilheight}{h}
\newcommand{\foillength}{\ell_z}
\newcommand{\fillfactor}{\lambda}

\newcommand{\lcs}[1]{\hat{#1}}
\newcommand{\fwmap}{f}
\newcommand{\locala}{\alpha}
\newcommand{\localb}{\beta}
\newcommand{\localc}{\gamma}

\newcommand{\domainlocala}{L_{\locala}}
\newcommand{\domainlocalb}{L_{\localb}}
\newcommand{\domainlocalc}{L_{\localc}}
\newcommand{\fwdomain}{\Omega_{\mathrm{fw}}}
\newcommand{\reffwdomain}{\lcs{\Omega}_{\mathrm{fw}}}

\newcommand{\foilcut}{\Sigma}
\newcommand{\reffoilcut}{\lcs{\foilcut}}

\newcommand{\normalvector}{\vec{n}}

\newcommand{\volfun}{\Phi}
\newcommand{\volfunlocal}{\hat{\volfun}}

\newcommand{\volfunFEmat}{\mathbf{u}}

\newcommand{\windingfun}{\vec{\zeta}}

\newcommand{\basisfun}{g}

\newcommand{\numbernodal}{N_{\mathrm{v}}}
\newcommand{\numberedge}{N_{\mathrm{w}}}
\newcommand{\numbervoltagefunction}{N_{\mathrm{g}}}
\newcommand{\turns}{N}
\newcommand{\numberelements}{N_{\mathrm{e}}}

\newcommand{\landau}{\mathcal{O}}

\newcommand{\voltage}{V}

\newcommand{\current}{I}
\newcommand{\volsource}{V_\mathrm{s}}

\newcommand{\impedanceLast}{R_\mathrm{L}}
\newcommand{\impedanceOne}{R }

\newcommand{\domainFW}{\Omega_\mathrm{fw}}

\newcommand{\lrBrackets}[1]{\left( #1 \right)}
\newcommand{\lrbrackets}[1]{\left[ #1 \right]}

\newcommand{\timestepth}{\Delta t_\mathrm{th}}
\newcommand{\timestepmg}{\Delta t_\mathrm{mg}}

\newcommand{\diff}{\mathrm{d}}
\newcommand{\absolutet}{\frac{\diff}{\diff t}}

\newcommand{\intV}[1]{\int_{\Omega}#1 \intsep \diff \Omega}

\newcommand{\sumNwj}{\sum_{j=1}^{\numberedge}}

\newcommand{\sumNvj}{\sum_{j=1}^{\numbernodal}}

\newcommand{\rt}{(\vec{r}, t)}
\newcommand{\rvec}{\vec{r}}

\DeclareMathOperator{\curl}{curl}
\DeclareMathOperator{\Div}{div}
\DeclareMathOperator{\grad}{grad}

\newcommand{\cmat}{\mathbf{c}}

\newcommand{\Gmat}{\mathbf{G}}
\newcommand{\Kmat}{\mathbf{K}}		
		
\newcommand{\Mmat}{\mathbf{M}}

\newcommand{\Xmat}{\mathbf{X}}

\DeclareAcronym{fe}{short={FE}, long={finite element}, list={Finite element}}
\DeclareAcronym{mqs}{short={MQS}, long={magnetoquasistatic}, list={Magnetoquasistatic}}
\DeclareAcronym{mvp}{short={MVP}, long={magnetic vector potential}, list={Magnetic vector potential}}
\DeclareAcronym{bc}{short={BC}, long={boundary condition}, list={Boundary condition}}
\DeclareAcronym{hce}{short={HCE}, long={heat conduction equation}, list={Heat conduction equation}}
\DeclareAcronym{dc}{short={DC}, long={direct current}, list={Direct current}}

\usepackage[bibstyle=ieee,citestyle=numeric-comp,maxnames=1, minnames=1,nohashothers=true,isbn=false,doi=true,url=false,sorting=none]{biblatex}
\usepackage{biblatex-shortfields}
\begin{document}
\title{Magneto-thermally Coupled Field Simulation of Homogenized Foil Winding Models}
\author{Silas Weinert, Jonas Bundschuh, Yvonne Späck-Leigsnering, and Herbert De Gersem
\thanks{Manuscript created in March, 2025, and revised in October, 2025; 
	The work is supported by the German Science Foundation (DFG project 436819664), the joint DFG/FWF Collaborative Research Centre CREATOR
	(DFG: Project-ID 492661287/TRR 361; FWF: 10.55776/F90, subprojects A03 and A04) at the Technical University of Darmstadt (TU Darmstadt), TU Graz and JKU Linz, the Graduate School Computational Engineering at TU Darmstadt, and the Athene Young Investigator Fellowship of the TU Darmstadt.
	
	All authors are with the Institute for Accelerator Science and Electromagnetic Fields at the Technical University of Darmstadt, 64289 Darmstadt, Germany. Jonas Bundschuh, Yvonne Späck-Leigsnering and Herbert De Gersem are also with the Graduate School of Excellence Computational Engineering at the Technical University of Darmstadt, 64289 Darmstadt, Germany. Corresponding author: Jonas Bundschuh. E-mail: \href{mailto:silas.weinert@stud.tu-darmstadt.de}{\color{black}{silas.weinert@stud.tu-darmstadt.de}}; \href{mailto:jonas.bundschuh@tu-darmstadt.de}{\color{black}{jonas.bundschuh@tu-darmstadt.de}};
	\href{mailto:spaeck@temf.tu-darmstadt.de}{\color{black}{spaeck@temf.tu-darmstadt.de}};  \href{mailto:degersem@temf.tu-darmstadt.de}{\color{black}{degersem@temf.tu-darmstadt.de}}}
}

\maketitle
\thispagestyle{fancy}

\begin{abstract}
Foil windings have, due to their layered structure, different properties than conventional wire windings, which make them advantageous for high frequency applications.
Both electromagnetic and thermal analyses are relevant for foil windings.
These two physical areas are coupled through Joule losses and temperature dependent material properties.
For an efficient simulation of foil windings, homogenization techniques are used to avoid resolving the single turns.
Therefore, this paper comprises a coupled magneto-thermal simulation that uses a homogenization method in the electromagnetic and thermal part.
A weak coupling with different time step sizes for both parts is presented.
The method is verified on a simple geometry and showcased for a pot transformer that uses a foil and a wire winding.
\end{abstract}

\begin{IEEEkeywords}
Foil windings, electromagnetic and thermal homogenization, eddy currents, finite element method
\end{IEEEkeywords}

\section{Introduction}
\IEEEPARstart{B}{esides} traditional wire or litz wire windings, foil windings are an alternative that are used in various application areas, for example in inductors \cite{Kazimierczuk_2010aa} or in transformers \cite{Sippola_2002a, Kurita_2023a, Gradinger_2021a} that are used in DC-DC converters \cite{CalderonLopez2019a, Ismail2023a}.
As the name suggests, a foil winding is built by winding a foil instead of a wire. 
This gives foil windings unique properties.
Compared to wire windings, they are easier and cheaper to construct \cite{Rios_2020aa,Barrios_2015aa}, have better mechanical properties \cite{Kazimierczuk_2010aa} and exhibit a lower \ac{dc} resistance \cite{Sullivan_2014aa}.
Furthermore, foil windings have better thermal properties \cite{Das_2020aa,Leuenberger_2015aa} because the foil acts as a highly thermally conductive connection from the interior to the surface and thus allows an efficient cooling. 
In contrast to that, the path from the interior to the surface is thermally insulated in wire windings, which makes cooling more difficult.

In the design of applications, not only the electromagnetic behavior but also the thermal behavior of the foil winding is of importance \cite{Sabariego_2024a}.
Both field aspects can be investigated with moderate effort using field simulation. 
Because of the layered structure of a foil winding with its very thin turns, a brute force simulation approach leads to unmanageably large models and long simulation times \cite{De-Gersem_2001aa}.
To mitigate this, there exist special homogenization models for electromagnetic simulations that avoid the resolution of every single turn of the foil winding but rather resolve the foil winding as a whole \cite{De-Gersem_2001aa,Geuzaine_2001aa,Dular_2002aa,Bundschuh2024ac}.
For the thermal simulation, a similar homogenization approach has to be used to also avoid resolving the turns \cite{Sabariego_2024a}. 

The electromagnetic and the thermal simulations are, however, not independent from each other. 
The Joule losses from the electromagnetic simulation act as heat source for the thermal simulation, and the temperature influences the material parameters of the electromagnetic simulation.
For that reason, this paper presents a magneto-thermally coupled simulation of foil windings that uses a homogenization model for the foil windings in both the electromagnetic and the thermal problem. 

The paper is structured as follows.
In Sec.~\ref{sec:definitions}, a foil winding is briefly defined and its description is presented. 
The multi-physical problem is presented in Sec.~\ref{sec:multiphysicy_proplem_setting}.
This includes the description of the \ac{mqs} and the thermal problem, the homogenization of the materials in the foil winding and the weak coupling scheme. 
The obtained equations are discretized in Sec.~\ref{sec:FEM}.
In Sec.~\ref{sec:validation}, the method is verified for a simple geometry, and in Sec.~\ref{sec:example}, a more realistic example of a pot transformer is analyzed. 
Finally, in Sec.~\ref{sec:conclusion}, a conclusion is drawn.

\section{Definitions}\label{sec:definitions}
\begin{figure}%
	\centering
	\begin{subfigure}{0.6\linewidth}
		\centering
		\includegraphics{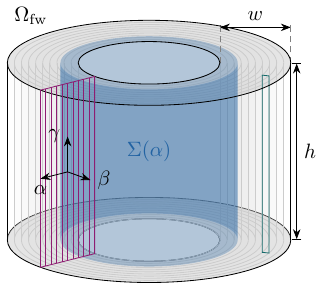}
		\caption{}
		\label{fig:FW Schematic}
	\end{subfigure}
	\begin{subfigure}{0.36\linewidth}
		\centering
		\includegraphics[]{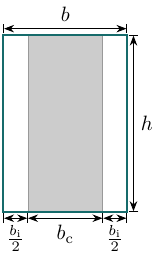}
		\caption{}
		\label{fig:FW Cross Section}
	\end{subfigure}
	\caption{Schematic representation of a foil winding. \subref{fig:FW Schematic} shows the foil winding domain $\domainFW$ with the local coordinate system $(\locala,\localb,\localc)$.
		Its coordinates are directed perpendicular to the turns, in winding direction and in the direction towards the tips of the turns, respectively. The constant cross section is highlighted in \accentname and the surface $\foilcut(\locala)$ (see \eqref{eq:def_foil_cut}) is illustrated for a fixed $\locala$ in \accenttwoname. The cross section of one turn is highlighted in \accentthreename and shown in detail in \subref{fig:FW Cross Section}, with the conducting material in gray and the insulation material in white. (Both figures are adapted from \cite{Bundschuh2024ac}).}
	\label{fig:FW Schematic and Cross Section}
\end{figure}%
A schematic representation of a foil winding is shown in Fig.~\ref{fig:FW Schematic and Cross Section}.
It has the width $\foilwidth$, height $\foilheight$ and consists of $\turns$ turns. 
The foil consists of a conducting material and an insulation material.
The subscripts \enquote*{c} and \enquote*{i} are used to indicate the respective affiliation.
The width of each turn is $\turnwidth = \turnwidthc + \turnwidthi$ (see Fig.~\ref{fig:FW Cross Section}). 
The fill factor $\fillfactor = \frac{\turnwidthc}{b}$ describes the proportion of the conductor thickness to the total foil thickness.
The foil winding domain $\fwdomain$ is described with a reference domain $\reffwdomain:= \domainlocala\times\domainlocalb\times\domainlocalc$ built from the intervals $\domainlocala,\domainlocalb,\domainlocalc\subset\R$ for the local coordinates and the (invertible) \emph{foil winding mapping} $\fwmap:\reffwdomain\mapsto\fwdomain$ \cite{Bundschuh2024ac}.
This equips the foil winding domain with the local coordinate system $(\locala,\localb,\localc)$ (see Fig.~\ref{fig:FW Schematic}).
Its unit vectors $\vec{e}_\locala$, $\vec{e}_\localb$ and $\vec{e}_\localc$ are directed perpendicular to the turns, in winding direction and towards the tips of the foil winding, respectively.
Note that all quantities related to the reference domain are denoted with a hat. 
Transformations to and from the foil winding domain are straightforward with the foil winding mapping $\fwmap$.
We define a foil cut on the reference domain as
\begin{equation}\label{eq:def_foil_cut}
	\reffoilcut(\locala^*):=\setprop{(\locala^*,\localb,\localc)\in\reffwdomain}{\localb\in\domainlocalb,\localc\in\domainlocalc}.
\end{equation}
The foil cut in the foil winding domain is depicted in Fig.~\ref{fig:FW Schematic} in \accenttwoname.

Throughout the paper, it is assumed that the foil is significantly thinner than the skin depth $\skindepth = \sqrt{2 /(\omega \permeabilityconductor \conductivityconductor)}$, with the angular frequency $\angularfrequency = 2 \pi \frequency$, the frequency $\frequency$, the permeability $\permeabilityconductor$ and the conductivity $\conductivityconductor$ of the conducting material. 
This relation results from the derivation of the homogenized foil winding model \cite{Paakkunainen_2024aa,Bundschuh2024ac}, where it is assumed that the current density is constant over the thickness of the foil.
Here, it is used to determine the valid frequency range for a given foil of a certain thickness and material parameters. 
For example, for a copper foil of thickness \qty{0.1}{\milli\meter} with $\conductivityconductor=\qty{60}{\mega\siemens\per\meter}$ and $\permeabilityconductor=\unit{\muzero}$ at $\frequency=\qty{5}{\kilo\hertz}$ the skin depth is about \qty{0.92}{\milli\meter}, which is significantly greater than the foil thickness.
Furthermore, it is assumed that the height of the foil winding is much larger than the width of one turn, i.e., $\foilheight \gg \turnwidth$.

\section{Multi-physics problem setting}\label{sec:multiphysicy_proplem_setting}

\subsection{Magnetoquasistatic Sub-Problem}\label{ssec:mqs_subproblem}
The $\mvp$-$\scalarpot$-formulation of the \ac{mqs} subset of Maxwell's equations is used, with the \ac{mvp} $\mvp\rt$ and electric scalar potential $\scalarpot\rt$.
The formulation reads
\begin{equation}
	\curl \lrBrackets{\reluctivity \curl \mvp } + \conductivity \partial_t \mvp + \sigma \grad \scalarpot = \sourcecurrentdensity,
\end{equation}
with the electric conductivity $\conductivity$, the reluctivity $\reluctivity$ and the source current density $\sourcecurrentdensity$. 

The homogenized foil winding model was introduced in \cite{De-Gersem_2001aa,Geuzaine_2001aa,Dular_2002aa}. 
Here, we use the version derived in \cite{Paakkunainen_2024aa,Bundschuh2024ac}. 
Therein, the voltage drop along the foil winding that is given by an external circuit is described with the \emph{voltage function} $\volfun(\rvec,t)$ that depends only on $\locala$ in the local coordinate system and on time, i.e., $\volfunlocal = \volfunlocal(\locala, t)$.
Multiplied with a \emph{distribution function} $\windingfun(\rvec)$ \cite{Schops_2013aa}, the relation $- \grad \scalarpot = \volfun \windingfun$ holds. 
The $\mvp$-$\scalarpot$-formulation with the homogenized foil winding model on the computational domain $\Omega\subseteq\R^3$ is given as \cite{Bundschuh2024ac}
\begin{subequations}
	\label{eq:MQS Homogenized}
	\begin{align}
		\curl \lrBrackets{\reluctivity \curl \mvp } + \conductivity \partial_t \mvp - \conductivity \volfun\windingfun &= \sourcecurrentdensity && \text{in } \Omega ,\label{eq:mqs_hom}\\
		\int_{\foilcut(\locala)} \conductivity \left(-\partial_t \mvp + \volfun \windingfun\right) \cdot \windingfun \intsep \dS &= \frac{\current}{\turnwidth} && \locala\text{ in }\domainlocala, \label{eq:Current Condition}
	\end{align}
\end{subequations}
together with appropriate initial and \acp{bc}.
The initial conditions must be consistent to the formulation and to the external circuit.
As \acp{bc}, the tangential components of the magnetic vector potential are forced to zero, which ensures that all magnetic flux is confined within the field model part and thus no further voltages are induced in the external circuit part.

The model ensures that the current $\current$ flows through each individual turn of the foil winding.
The voltage drop over the foil winding is given by \cite{Bundschuh2024ac}
\begin{equation}\label{eq:voltage_function}
	\voltage = \frac{1}{\turnwidth}\int_{\domainlocala} \volfunlocal(\locala)\intsep\diff\locala.
\end{equation}

\subsection{Thermal Sub-Problem}
Heat transfer is described by the transient \ac{hce} as \cite{VDI10Atlas}
\begin{equation}\label{eq:Heat Conduction Equation}
	- \Div \left(\thermalconductivity \grad \temperature\right) + \cV \partial_t \temperature = \heatsource,
\end{equation}
with the temperature $\temperature$, the volumetric heat capacity $\cV$, the thermal conductivity $\thermalconductivity$, and the volumetric heat source $\heatsource$. 
The heat dissipation through the boundary can be described by the Robin \ac{bc}
\begin{equation}
	\thermalconductivity \normalvector\cdot\grad T + \thConv (\temperature - \temperatureAmb) = 0,
\end{equation}
where $\normalvector$ is the outward pointing normal vector at the boundary, $\thConv$ is the convection coefficient, and $\temperatureAmb$ is the reference ambient temperature. 
The limits $\thConv \rightarrow \infty$ and $\thConv = 0$ correspond to a homogeneous Dirichlet and Neumann \ac{bc}, respectively.

\subsection{Material Homogenization}
The homogenized foil winding model does not resolve the single turns in the foil winding domain. 
Instead, the foil winding domain is replaced with an artificial homogenized material.
Its material properties are determined from the properties and dimensions of the conducting and insulating material.
With respect to the local coordinate system, the material properties are diagonal tensors \cite{Bundschuh2024ac}, i.e.,
\begin{equation}
	\lcs{m} = \begin{pmatrix}
		m_\perp &0 & 0\\
		0& m_\parallel & 0\\
		0& 0& m_\parallel
	\end{pmatrix},
\end{equation}
for a general material property $\lcs{m}$, where $m\in\left\{\reluctivity,\conductivity,\thermalconductivity\right\}$. 
The material property perpendicular to the turns is denoted $m_\perp$ and parallel to the turns $m_\parallel$.
Inside the foil winding domain the material properties are homogenized by mixing rules \cite{Sihvola_1999aa,VDI10Atlas} as 
\begin{subequations}
	\begin{align}
		\reluctivity_\perp &= \fillfactor \reluctivityconductor + (1 - \fillfactor) \reluctivityinsulator, & \frac{1}{\reluctivity_\parallel} &= \frac{\fillfactor}{\reluctivityconductor} + \frac{1-\fillfactor}{\reluctivityinsulator},\\
		\frac{1} {\conductivity_\perp}&= \frac{\fillfactor}{\conductivityconductor} + \frac{1-\fillfactor}{\conductivityinsulator}, & \sigma_\parallel &= \fillfactor \conductivityconductor + (1 - \fillfactor)\conductivityinsulator,\\
		\frac{1}{\thermalconductivity_\perp}&= \frac{\fillfactor}{\thermalconductivityconductor} + \frac{1-\fillfactor}{\thermalconductivityinsulator}, & \thermalconductivity_\parallel &= \fillfactor \thermalconductivityconductor + (1 - \fillfactor)\thermalconductivityinsulator.
	\end{align}
\end{subequations}
Due to the vanishing electric conductivity of the insulation material ($\conductivityinsulator = 0$), the electric conductivity perpendicular to the turns is also zero, i.e., $\sigma_\perp = 0$. In addition, the volumetric heat capacity is homogenized using the relative volume ratios between the materials \cite{Driesen_2000aa}.

\subsection{Coupled Formulation}
In the following, the thermal model and the magnetic model are mutually coupled.
The Joule losses $\heatsource = \vec{J} \cdot \vec{E}$ from the magnetic model are the heat source for the thermal model, and the temperature $\temperature$ from the thermal model influences the electric conductivity $\conductivity(\temperature)$ of the magnetic model.
With the homogenized foil winding model from \eqref{eq:MQS Homogenized}, the Joule losses are
\begin{equation}\label{eq:Coupling Condition}
	\heatsource = \lrBrackets{- \partial_t \mvp + \volfun \windingfun} \cdot \lrBrackets{-\sigma \partial_t \mvp + \sigma \volfun \windingfun}.
\end{equation}
In Section \ref{sec:FEM}, this term is averaged for every finite element.
For the electric conductivity, the expression
\begin{equation}
	\sigma(\temperature) = \frac{\conductivityRef}{1 + \temperaturecoefficientRef (\temperature - \temperatureRef)}
\end{equation}
is used, where $\temperaturecoefficientRef = \qty[per-mode=power]{3.93e-3}{\per\kelvin}$ is the temperature coefficient and $\conductivityRef = \qty{60}{\mega\siemens\per\meter}$ the material's electric conductivity at reference temperature $\temperatureRef=\qty{293.15}{\kelvin}$ \cite{Driesen_2000aa, DAngelo_2017aa}.

\begin{figure}%
	\centering
	\includegraphics[]{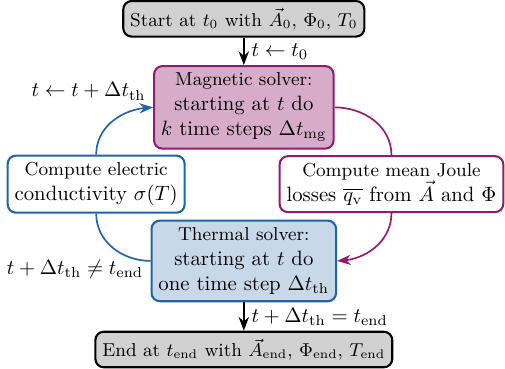}
	\caption{Weak coupling scheme. It holds $k\timestepmg=\timestepth$ with $k\in\N$.}
	\label{fig:Weak Coupling Scheme}
\end{figure}

The coupled problem is solved using an iterative method \cite{Hameyer1999aa}, as illustrated in Fig.~\ref{fig:Weak Coupling Scheme}.
After initializing the iteration with the initial conditions for the fields, the magnetic and thermal solvers are executed sequentially.
The magnetic solver performs several magnetic time steps with a fixed temperature distribution. 
Then, the Joule losses are computed and averaged over the elapsed time. 
With this information, the thermal solver performs one thermal time step which yields the temperature distribution needed for the electric conductivity in the magnetic solver.
By that, the thermal problem can be solved with a much coarser time step than the magnetic problem.
The iteration is continued until the final time is reached.

\section{Finite-Element Discretization} \label{sec:FEM}
The \ac{mvp} $\mvp\rt$ is discretized with $\numberedge$ lowest order \ac{fe} edge shape functions $\vec{w}_j(\rvec)$, i.e., Nédélec elements \cite{Nedelec1980a}, and the temperature $\temperature\rt$ is discretized with $\numbernodal$ lowest order \ac{fe} nodal shape functions $v_j(\rvec)$, i.e., Lagrange elements \cite{Bossavit_1998aa}.
The approximations read
\begin{subequations}
	\label{eq:Discretization MVP and Temperature}
	\begin{align}
		\mvp\rt &\approx \sumNwj \mvpFE_j(t) \vec{w}_j(\vec{r}),\\
		\temperature\rt &\approx \sumNvj \temperatureFE_j(t) v_j(\vec{r}).
	\end{align}
\end{subequations}
The voltage function $\volfun(\rvec, t)$ is discretized with $\numbervoltagefunction$ basis functions $\basisfun_j(\vec{r})$ that are tailored to the foil winding domain. The approximation reads
\begin{equation}
	\volfun\rt \approx \sum_{j=1}^{\numbervoltagefunction} u_j(t) \basisfun_j(\vec{r}).
\end{equation}
The basis functions $\basisfun_j(\vec{r})$ are constant in $\vec{e}_\localb$- and $\vec{e}_\localc$-direction and vary only in $\vec{e}_\locala$-direction.
Thus, they are easily defined on the reference foil winding domain \cite{Paakkunainen_2024aa}.

The Ritz-Galerkin method is applied to \eqref{eq:MQS Homogenized} and \eqref{eq:Heat Conduction Equation}.
For the magnetic sub-problem, this defines matrices $\Kmat_\nu,\Mmat_\sigma \in \R^{\numberedge\times\numberedge}$, $\Xmat_\sigma\in\R^{\numberedge\times\numbervoltagefunction}$, $\Gmat_\sigma\in\R^{\numbervoltagefunction\times\numbervoltagefunction}$ and vectors $\qmatmg\in\R^{\numberedge}$, $\mathbf{c}\in\R^{\numbervoltagefunction}$.
For the thermal sub-problem, this defines matrices $\Kmat_\thermalconductivity,\Mmat_{\cV} \in \R^{\numbernodal\times\numbernodal}$ and the vector $\qmatth \in \R^{\numbernodal}$.
Their entries are defined by
\begin{subequations}
	\begin{align}\label{eq:MQS Matrices}
		\lrbrackets{\Kmat_\nu}_{i,j}			&= \intV{\nu \curl \vec{w}_j \cdot \curl \vec{w}_i}	,\\
		\lrbrackets{\Mmat_\sigma}_{i,j} 		&= \intV{\sigma \vec{w}_j \cdot \vec{w}_i}	,\\
		\lrbrackets{\Xmat_\sigma}_{i,j}		&= \intV{\sigma \basisfun_j \windingfun \cdot \vec{w}_i} ,\\
		\lrbrackets{\Gmat_\sigma}_{i,j}		&= \intV{\sigma \windingfun \cdot \windingfun \basisfun_j \basisfun_i},	\\
		\lrbrackets{\qmatmg}_i				&= \intV{\vec{J}_{\mathrm{s}} \cdot \vec{w}_i},	\\
		\lrbrackets{\mathbf{c}}_i			&= \frac{1}{b} \int_{L_\locala} \hat{\basisfun}_i \intsep \diff \locala,
	\end{align}
\end{subequations}
and 
\begin{subequations}
	\begin{align}
		\lrbrackets{\Kmat_\thermalconductivity}_{i,j}		&= \intV{\thermalconductivity \grad v_j \cdot \grad v_i}, \\
		\lrbrackets{\Mmat_{\cV}}_{i,j} 		&= \intV{\cV v_j v_i},\\
		\lrbrackets{\qmatth}_i				&= \intV{\heatsource v_i}.
	\end{align}
	\label{eqs:Thermal Matrices}
\end{subequations}

The degrees of freedom are collected in vectors $\mvpFEmat=\left[a_1,\dots,a_{\numberedge}\right]\transpose$, $\volfunFEmat=\left[u_1,\dots,u_{\numbervoltagefunction}\right]\transpose$ and $\temperaturemat=\left[\temperatureFE_1,\dots,\temperatureFE_{\numbernodal}\right]\transpose$. The semi-discrete system of equations from \eqref{eq:MQS Homogenized}, \eqref{eq:voltage_function} and \eqref{eq:Heat Conduction Equation} reads
\begin{subequations}
	\label{eq:FEM Semi-Discrete System}
	\begin{align}
		\Kmat_\nu \mvpFEmat + \Mmat_\sigma \absolutet \mvpFEmat - \Xmat_\sigma \volfunFEmat &= \qmatmg,\\
		- \Xmat_\sigma^\top \absolutet \mvpFEmat + \Gmat_\sigma \volfunFEmat - \cmat \current &= 0,\\
		\cmat^\top \volfunFEmat - \voltage &= 0 ,\\
		\Kmat_\thermalconductivity \temperaturemat + \Mmat_{\cV} \absolutet \temperaturemat &= \qmatth.
	\end{align}
\end{subequations}
The system is completed with appropriate initial, circuit and boundary conditions (see Sec.~\ref{ssec:mqs_subproblem}).
The backward Euler method is used for time discretization.

\section{Verification}\label{sec:validation}
We verify the results of \eqref{eq:FEM Semi-Discrete System} against a reference solution computed with \ComsolMultiphysics \cite{Comsol}.
The simulations of the coupled problem are carried out using \Pyrit, a \ac{fe} solver in \Python \cite{Bundschuh_2023ab}.

\begin{figure}%
	\centering
	\begin{subfigure}{0.48\linewidth}
		\centering
		\includegraphics[]{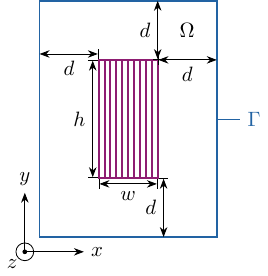}
	\end{subfigure}
	\begin{subfigure}{0.48\linewidth}
		\centering
		\SetTblrInner{rowsep=1pt,colsep=4pt} 
		\begin{tblr}{columns={font=\small},
				column{1}={l},
				column{2,3}={c},
				hline{1,2,Z}={tablelinecolor},
				row{odd} = {bg=tablezebracolor}, row{1} = {bg=white},
			}
			Parameter & Value\\ 
			$\foilwidth$ & $\SI{10}{\milli\meter}$ \\
			$\foilheight$ & $\SI{20}{\milli\meter}$ \\
			$\foillength$ & $\SI{50}{\milli\meter}$ \\
			$d$ & $\SI{10}{\milli\meter}$\\
			$\turns$ & $\num{30}$ \\
			$\fillfactor$ & $\qty{80}{\percent}$\\
		\end{tblr}
		\vspace{0.65cm}
	\end{subfigure}
	\caption{Geometry of the verification model with the dimensions in the table. The foil winding (\accentname) is centered within a box filled with air. On the boundary $\domainboundary$, an electric \ac{bc} and an isothermal \ac{bc} is applied to the magnetic and thermal problem, respectively. (Adapted from \cite{Bundschuh2024ac}).}
	\label{fig:Winding In Air}
\end{figure}

The considered model is shown in Fig.~\ref{fig:Winding In Air}, and the material parameters are collected in Tab.~\ref{tab:Materials for Validation}.
The foil winding is placed in the center of a box filled with air.
On the boundary $\domainboundary$, a homogeneous electric \ac{bc} is used for the magnetic sub-problem, and an isothermal \ac{bc} at $\temperatureAmb$ is used for the thermal sub-problem.
The magnetic simulation is carried out using peak values in the frequency domain at $\frequency = \qty{50}{\hertz}$, and the thermal simulation in the time domain using a constant time step of $\timestepth = \qty{2}{\minute}$ over a period of $\qty{10}{\hour}$.
In the frequency domain, the average heat losses are
\begin{equation}\label{eq:Coupling Condition Harmonic}
	\overline\heatsource =\frac{1}{2} \lrBrackets{- \jomega \underline\mvp + \underline\volfun \windingfun} \cdot \lrBrackets{-\sigma \jomega \underline\mvp + \sigma \underline\volfun \windingfun}^*,
\end{equation}
where $^*$ denotes the complex conjugate and the underscore indicates that the fields are complex valued. 
The voltage function is discretized with $\numbervoltagefunction = 7$ basis functions.
\begin{table}%
	\caption{Material properties used for the verification model and the example. The values are based on \cite{VDI10Atlas}.}
	\label{tab:Materials for Validation}
	\begin{tblr}{
			column{1}={l},
			column{2,3}={c},
			hline{1,2,Z}={tablelinecolor},
			row{odd} = {bg=tablezebracolor}, row{1} = {bg=white},
			cell{2-5}{2-5}={r},
			cell{1}{2-5}={c},}
		Material & $\conductivity$ & $\permeability$ & $\thermalconductivity$ & $\volheatcap$\\
		Conductor & \qty{60}{\mega\siemens\per\meter} & $\qty{}{\muzero}$ & \qty{385}{\watt\per\meter\per\kelvin} & \qty{3.45}{\mega\joule\per\meter\cubed\per\kelvin} \\
		Insulator & \qty{0}{\siemens\per\meter} & $\qty{}{\muzero}$ & \qty{90}{\milli\watt\per\meter\per\kelvin} & \qty{1.03}{\mega\joule\per\meter\cubed\per\kelvin} \\
		Iron & \qty{0}{\siemens\per\meter} & $\qty{5000}{\muzero}$ & \qty{72}{\watt\per\meter\per\kelvin} & \qty{3.53}{\mega\joule\per\meter\cubed\per\kelvin} \\
		Air & \qty{0}{\siemens\per\meter} & $\qty{}{\muzero}$ & \qty{26}{\milli\watt\per\meter\per\kelvin} & \qty{1}{\kilo\joule\per\meter\cubed\per\kelvin} \\
	\end{tblr}
\end{table}%

\begin{figure}%
	\centering
	\includegraphics[]{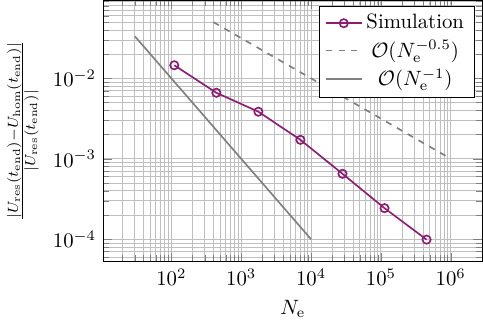}
	\caption{Relative error between the resolved thermal internal energy $\energythres$ and the homogenized thermal inner energy $\energythhom$ at the time instance $t_\text{end}$ over the number of mesh elements $\numberelements$.}
	\label{fig:Convergence thermal energy}
\end{figure}

Figure~\ref{fig:Convergence thermal energy} shows the relative error of the thermal internal energy 
\begin{equation}
	U = \intV{ \volheatcap \temperature }
\end{equation}
between the resolved and homogenized results over the number of mesh elements $\numberelements$.
The relative error is shown for the last time step of the simulation.
The results show a convergence order between $\landau(\numberelements^{-0.5})$ and $\landau(\numberelements^{-1})$.

\section{Example: Pot Transformer}\label{sec:example}
\begin{figure}%
	\centering
	\begin{subfigure}{0.58\linewidth}
		\centering
		\includegraphics[]{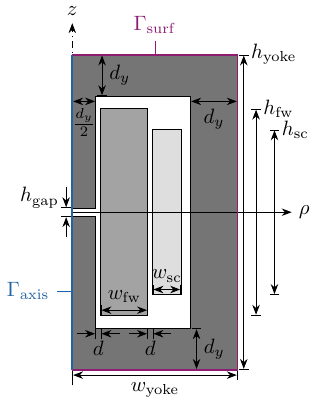}
	\end{subfigure}
	\begin{subfigure}{0.4\linewidth}
		\centering
		\SetTblrInner{rowsep=1pt,colsep=4pt} 
		\begin{tblr}{columns={font=\small},
				column{1,2}={c},
				hline{1,2,Z}={tablelinecolor},
				row{odd} = {bg=tablezebracolor}, row{1} = {bg=white},
				cell{2-Z}{2}={r},
			}
			Parameter & Value \\ 
			$w_\text{yoke}$ & $\qty{35}{\milli\meter}$ \\
			$h_\text{yoke}$ & $\qty{76.2}{\milli\meter}$ \\
			$w_\text{fw}$ & $\qty{10}{\milli\meter}$ \\
			$h_\text{fw}$ & $\qty{50}{\milli\meter}$ \\
			$w_\text{sc}$ & $\qty{6}{\milli\meter}$ \\
			$h_\text{sc}$ & $\qty{40}{\milli\meter}$ \\
			$d_y$ & $\qty{10}{\milli\meter}$ \\
			$d$ & $\qty{1}{\milli\meter}$ \\
			$h_\text{gap}$ & $\qty{2}{\milli\meter}$ \\	
		\end{tblr}
		\vspace{0.45cm}
	\end{subfigure}
	\caption{Geometry and dimensions of the pot transformer. The inner foil winding and the outer wire winding (both in light gray) are surrounded by the yoke (in dark gray). The yoke has an air gap in the center limb and is filled with air (white).}
	\label{fig:Pot Transformer}
\end{figure}

\begin{figure}%
	\centering
	\includegraphics[]{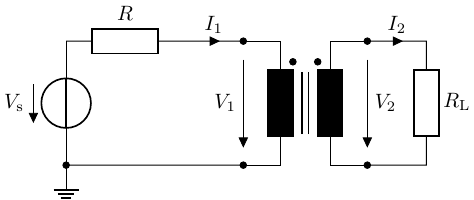}
	\caption{Circuit including the pot transformer. The values of the components are $\impedanceOne = \qty{1}{\ohm}$, $\impedanceLast = \qty{200}{\ohm}$, $\volsource = \qty{50}{\volt}$, $\frequency = \qty{5}{\kilo\hertz}$.}
	\label{fig:PT Circuit}
\end{figure}

As a more realistic example, we consider a pot transformer. 
Its geometry and dimensions are shown in Fig.~\ref{fig:Pot Transformer}.
The primary winding is a foil winding with $\turns = 100$ turns and a fill factor of $\fillfactor_1 = \qty{80}{\percent}$, and the secondary winding is a wire winding with $\turns = 500$ turns and a fill factor of $\fillfactor_2 = \qty{80}{\percent}$.
The transformer is connected to a small circuit, consisting of a voltage source and two resistors, as shown in Fig.~\ref{fig:PT Circuit}.
A monolithic coupling to the circuit is used within \Pyrit.

The voltage source uses a sinusoidal voltage with an amplitude of $\qty{50}{\volt}$ and a frequency of $\frequency=\qty{5}{\kilo\hertz}$.
On the boundary $\domainboundary_{\mathrm{surf}}\cup\domainboundary_{\mathrm{axis}}$, a homogeneous electric \ac{bc} is used for the magnetic sub-problem.
For the thermal sub-problem, a homogeneous Neumann \ac{bc} is applied on $\domainboundary_{\mathrm{axis}}$, while a Robin \ac{bc} with a convection coefficient $\thConv = \qty{25}{\watt\per\meter\squared\per\kelvin}$ is applied on $\domainboundary_{\mathrm{surf}}$.
For this example, both the magnetic and the thermal sub-problem are solved in time domain on the interval $\left[\qty{0}{\hour}, \qty{10}{\hour}\right]$. 
The magnetic sub-problem uses \num{200} time steps per period of the exciting voltage, i.e., $\timestepmg=\qty{1}{\micro\second}$, and the thermal sub-problem uses an initial time step of $\timestepth=\qty{30}{\second}$ and is increased when approaching the steady state temperature (see Tab.~\ref{tab:thermaltimesteps}).
\begin{table}
	\centering
	\caption{Thermal time steps $\timestepth$ on different time intervals for the pot transformer example.}
	\label{tab:thermaltimesteps}
	\begin{tblr}{
		column{1,2}={l},
		hline{1,2,Z}={tablelinecolor},
		row{odd} = {bg=tablezebracolor}, row{1} = {bg=white},
		}
		Time interval & Thermal time step\\
		$\left[\qty{0}{\minute},\qty{10}{\minute}\right]$ & $\timestepth=\qty{30}{\second}$\\
		$\left[\qty{10}{\minute},\qty{1}{\hour}\right]$ & $\timestepth=\qty{2}{\minute}$\\
		$\left[\qty{1}{\hour},\qty{2}{\hour}\right]$ & $\timestepth=\qty{5}{\minute}$ \\
		$\left[\qty{2}{\hour},\qty{4}{\hour}\right]$ & $\timestepth=\qty{10}{\minute}$\\
		$\left[\qty{4}{\hour},\qty{10}{\hour}\right]$ & $\timestepth=\qty{20}{\minute}$ \\
	\end{tblr}
\end{table}

\begin{figure}%
	\centering
	\begin{subfigure}{0.41\linewidth}
		\centering
		\includegraphics[width=\linewidth]{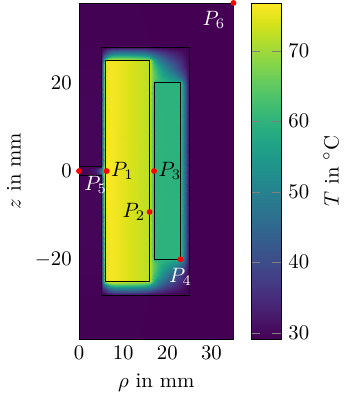}
		\caption{}
		\label{fig:PT Temperature Steady}
	\end{subfigure}
	\begin{subfigure}{0.56\linewidth}
		\centering
		\includegraphics[width=\linewidth]{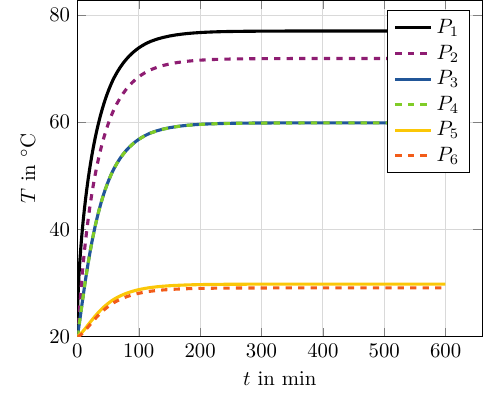}
		\caption{}
		\label{fig:PT Temperature Over Time}
	\end{subfigure}
	\caption{\subref{fig:PT Temperature Steady} shows the steady state temperature of the pot transformer, and \subref{fig:PT Temperature Over Time} shows the temperature rise over time for the points $P_1,\dots,P_6$ illustrated in \subref{fig:PT Temperature Steady}.}
	\label{fig:PT Temperature}
\end{figure}

\begin{figure}%
	\centering
	\begin{subfigure}{0.41\linewidth}
		\centering
		\includegraphics[width=\linewidth]{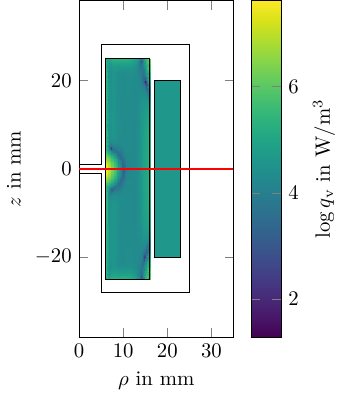}
		\caption{}
		\label{fig:PT Heat Source}
	\end{subfigure}
	\begin{subfigure}{0.56\linewidth}
		\centering
		\includegraphics[width=\linewidth]{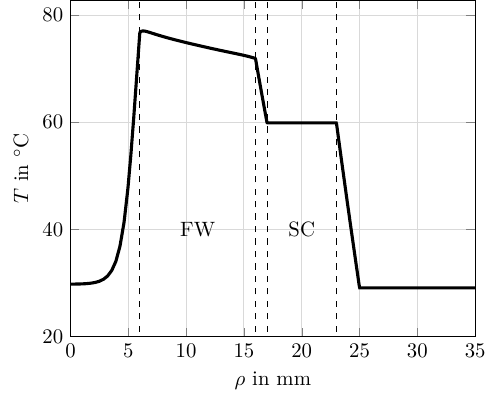}
		\caption{}
		\label{fig:PT Temperature over radius}
	\end{subfigure}
	\caption{\subref{fig:PT Heat Source} shows the heat losses of the pot transformer, and \subref{fig:PT Temperature over radius} shows the temperature distribution along the red line in \subref{fig:PT Heat Source}, i.e., along the radial coordinate $\rho$ for at $z = 0$, marking the foil winding (FW) and the wire winding (SC).}
	\label{fig:PT Temperature Steady and Radius}
\end{figure}

The steady state temperature distribution of the pot transformer is shown in Fig.~\ref{fig:PT Temperature Steady}.
Both windings have a relatively even temperature distribution, where the foil winding has a temperature of $\qty{77}{\degreeCelsius}$ and the wire winding of $\qty{60}{\degreeCelsius}$. 
Figure~\ref{fig:PT Temperature Steady} also defines six points at which the temperature rise over time is visualized in Fig.~\ref{fig:PT Temperature Over Time}.
Here, small temperature differences in the foil and wire winding are visible.
In Fig.~\ref{fig:PT Heat Source}, the heat losses are plotted in the pot transformer. 
The losses are evenly distributed in the wire winding because eddy currents are disregarded here. 
In the foil winding, we can observe variations in the losses with peak values close to the air gap where high eddy currents are induced by fringing fluxes.
Finally, Fig.~\ref{fig:PT Temperature over radius} shows the temperature distribution on a radial line at $z=0$.
Here, we can see again the highest temperature in the foil winding near the air gap and also the even temperature in the wire winding.

\section{Conclusion}\label{sec:conclusion}
This paper presented a coupled magneto-thermal simulation procedure for analyzing foil windings, accounting for both electromagnetic and thermal effects. 
Due to the interdependence between both phenomena, a weakly coupled approach was employed.
This enables the use of separate solvers with different time step sizes for the \ac{mqs} and thermal sub-problems.
To reduce computational effort, a homogenization model was applied for both sub-problems, eliminating the need to resolve individual foil layers in the mesh.
The proposed method was successfully verified against results obtained from \ComsolMultiphysics.
As a practical application, a pot-type transformer with an external electrical circuit was investigated.
The analysis identified a thermal hot spot in the foil winding near the air gap of the yoke, demonstrating the method's capability to predict critical temperature distributions in realistic configurations.

\printbibliography

\appendix\label{sec:appendix}
\section{Errata}
In the published version (\textit{Advanced Modeling and Simulation in Engineering Sciences}, vol.~12, no.~1, Oct.~2025, \textsc{DOI}: \href{https://doi.org/10.1186/s40323-025-00315-4}{10.1186/s40323-025-00315-4}), equations \eqref{eq:mqs_hom} and \eqref{eq:Current Condition} were renumbered as (3) and (4). Each citation of (3) on pages 5–7 should instead cite both (3) and (4).

\end{document}